\documentclass[fleqn,10pt]{wlscirep}
\pdfoutput=1

\title{Critical scaling of icosahedral medium-range order in CuZr metallic glass-forming liquids}

\author[1]{Z. W. Wu}
\author[2]{F. X. Li}
\author[2]{C. W. Huo}
\author[2,*]{M. Z. Li}
\author[3]{W. H. Wang}
\author[4]{K. X. Liu}
\affil[1]{International Center for Quantum Materials, School of Physics,
Peking University, Beijing 100871 China}
\affil[2]{Department of Physics, Renmin University of China, Beijing 100872 China}
\affil[3]{Institute of Physics, Chinese Academy of Sciences, Beijing 100190 China}
\affil[4]{Department of Mechanics and Engineering Science, Peking University, Beijing 100871 China}

\affil[*]{maozhili@ruc.edu.cn}

\begin{abstract}
The temperature evolution of icosahedral medium-range order formed by interpenetrating icosahedra in
CuZr metallic glass-forming liquids was investigated via molecular dynamics simulations. Scaling analysis based on
percolation theory was employed, and it is found that the size distribution of clusters formed by the central atoms
of icosahedra at various temperatures follows a very good scaling law with the cluster number density scaled
by $S^{-\tau}$ and the cluster size $S$ scaled by $\left|1-T_c/T\right|^{-1/\sigma}$, respectively. Here $T_c$ is
scaling crossover-temperature. $\tau$ and $\sigma$ are scaling exponents. The critical scaling behaviour suggests that
there would be a structural phase transition manifested by percolation of locally favoured structures
underlying the glass transition, if the liquid could be cooled slowly enough but without crystallization intervening.
Furthermore, it is revealed that when icosahedral short-range order (ISRO) extends to medium-range length scale by connection,
the atomic configurations of ISROs will be optimized from distorted ones towards more regular ones gradually,
which significantly lowers the energies of ISROs and introduces geometric frustration simultaneously.
Both factors make key impacts on the drastic dynamic slowdown of supercooled liquids.
Our findings provide direct structure-property relationship for understanding the nature of glass transition.
\end{abstract}
\begin{document}

\flushbottom
\maketitle

\thispagestyle{empty}

\section*{Introduction}

The nature of glass and glass transition is the deepest and most interesting unsolved problem
in the solid state science\cite{Anderson-sci,Debenedetti-nat,Berthier-rmp,Ediger-jcp}.
With a generic definition, numerous systems spanning a broad range of length scale such as
atomic and colloidal systems, foams, and granular materials, can be considered as glass
when certain conditions are satisfied\cite{Berthier-rmp}. Metallic glass, as a relatively ``simple'' glassy system
for scientific research of glass transition and promising industrial material\cite{Zhong-nat},
has attracted much attention and interest of scientists from broad research fields\cite{Cheng-pms,Wang-pms}.

The CuZr metallic glass-former has been extensively investigated
\cite{Cheng-prb,Li-prb,Wu-prb,Hao-prb,Wakeda-am,Li-sci,Soklaski-prb,Soklaski-pm,Ding-am}.
It exhibits good glass-forming ability, and its binary chemical composition
reduces the complexity of local atomic structures, which makes this system a good model
for the study of structure-property relationships in liquids and glasses. Previous studies demonstrate
that icosahedral short-range order (ISRO) is closely correlated with the slow dynamics
and dynamical heterogeneity during glass formation
\cite{Cheng-prb,Li-prb,Wu-prb,Hao-prb,Wakeda-am,Soklaski-prb,Soklaski-pm,Ding-am}.
It has also been demonstrated that the formation of icosahedral medium-range structures
via the interpenetration of ISROs plays a key role in the dynamic slowdown\cite{Wu-prb,Hao-prb},
The more the ISROs are connected, the slower the dynamics of the connected ISROs is.
This implies that the icosahedral medium-range order formed by the percolation of ISROs
may be related to the glass transition. So far, the concept of percolation of ISROs has been
widely used for understanding the relationship between structural evolution and glass
transition in metallic glass-forming liquids. However, the question is that no
specific percolation theory or scaling analysis has been derived to establish a
direct link between the percolation of ISROs and glass transition in the metallic
glass-forming liquids. In addition, the population of ISROs in CuZr metallic glass-forming
liquids sometimes is not very high so that ISROs do not even percolate as glass
transition occurs\cite{Soklaski-prb}, For example, Cu$_{50}$Zr$_{50}$ is a good glass-former in both
experiments\cite{Li-sci} and computer simulations\cite{Wu-nc,Tang-nm}.
However, the fraction of ISROs in Cu$_{50}$Zr$_{50}$ at 300 K is found to be less than 4\% in simulation\cite{Li-prb}.
Even in the inherent structure of Cu$_{50}$Zr$_{50}$, it is less than 15\%\cite{Cheng-pms}.
To establish the link between percolation of ISROs and glass transition,
the nearest-neighbor atoms of ISROs were also taken into account for the percolation.
Meanwhile, the percolation of ISROs in CuZr metallic glass-forming liquids
could be influenced by system size in simulation, leading to uncertainty for understanding
the strcuture-dynamics relationships in these systems. Therefore, it is highly desirable to develop a scaling
analysis based on percolation theory to establish a quantitative description of percolation
of ISROs and link to glass transition. Moreover, although ISROs are found to correlate
with the slow dynamics and the connection of ISROs makes it even slower, the physical
origin is still not very clear.

In this work, molecular dynamics simulations were performed for Cu$_{50}$Zr$_{50}$
with realistic interatomic potential (see Methods). We investigated the connection of
ISROs via interpenetrating or volume-sharing. Graph theory was introduced to characterize the the clusters formed by
the central atoms of interpenetrating ISROs at different temperatures as the CuZr
metallic glass-forming liquids are cooled down, and the \emph{equilibrium}
cluster size distribution was analyzed. Scaling analysis based on \emph{percolation theory}\cite{Christensen-book,Stauffer-book}
was conducted for the cluster size distribution. It is found that the cluster size distributions
at various temperatures collapse together and follow a good scaling law, as the cluster
size is $S$ scaled by $\left|1-T_c/T\right|^{-1/\sigma}$ and the cluster number density is scaled
by $S^{-\tau}$, respectively. The scaling analysis suggests that there could exist a geometric
phase transition of percolation of locally favoured structures, once the metallic glass-forming
liquids are quenched slowly enough (in the hypothetical limit of infinitely long relaxation time)
but without crystallization intervening, and glass transition
may be related to the percolation of locally favoured structures. Furthermore, it is revealed
that as ISROs are connected together, the atomic configurations of connected ISROs are
optimized towards more regular icosahedra. The optimized ISROs enhance the geometric
frustration, and the local energies are significantly lowered, which stabilizes the structure of
liquids and slows down the dynamics in glass transition.

\section*{Results and Discussions}

\subsection*{Scaling analysis for the percolation of icosahedral network}

\begin{figure}[ht]
\centering
\includegraphics[width=0.5\linewidth]{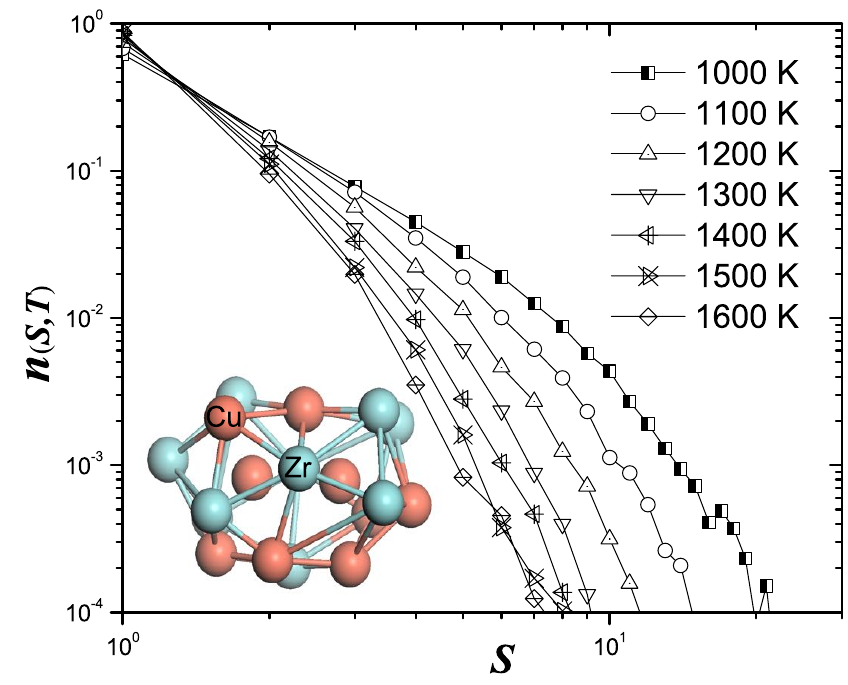}
\caption{Number density distributions of clusters formed by the connection of ISROs
via volume sharing at different temperatures during cooling. The inset illustrates a cluster
with size $S=2$ formed by two volume-shared ISROs.\label{sd}}
\end{figure}

\begin{figure}[ht]
\centering
\includegraphics[width=0.5\linewidth]{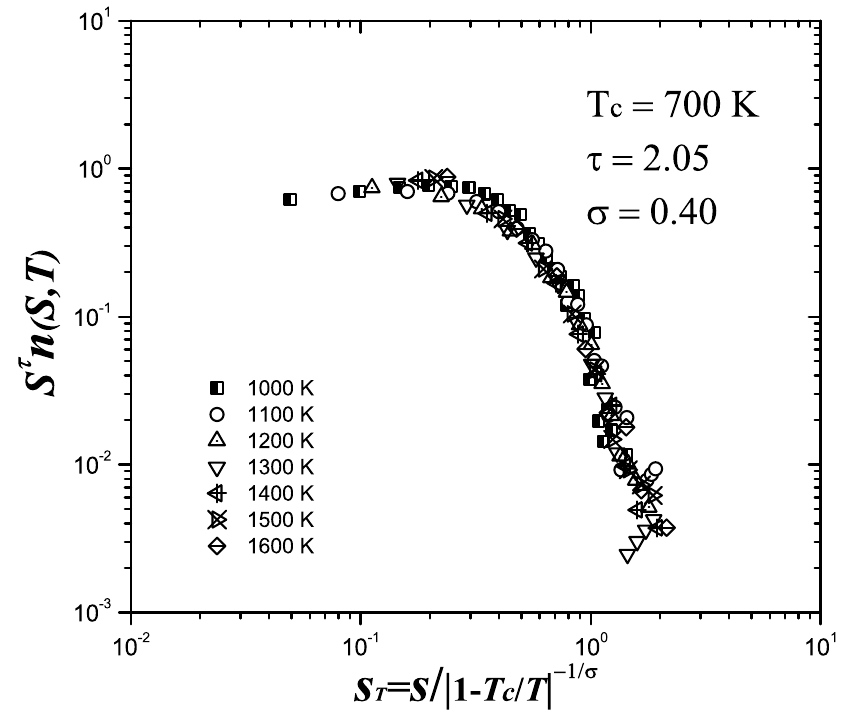}
\caption{The scaled size distribution follows an excellent scaling behaviour with the
scaled cluster size $S_T$. Here $S$ is the cluster size. $T_c$ is
the critical point. $\tau$ and $\sigma$ are scaling critical exponents.
It can be seen that all the cluster number densities fall onto the same curve
representing the graph of scaling function $f(S/S_c)$. \label{sl}}
\end{figure}

First, we analyzed the size distributions of clusters formed by the connection of central atoms
of the icosahedra at various temperatures. In our analysis, graph theory was applied for the construction of
connection of icosahedra and the resulting icosahedral network (see Methods).
As shown in Fig.~\ref{sd}, the size distributions decrease monotonically as cluster size increases
and exhibit similar behaviour at different temperatures. The size distributions in small size part follow
power-law behaviour, but deviate in larger size part, which signals there is a finite characteristic
cluster size in the system and, for a given temperature $T$, the characteristic cluster size marks the crossover
between a power-law behaviour and a rapid decay of $n(S,T)$, qualitatively.
As temperature decreases, more and more
larger clusters are formed in the supercooled liquids. It would be expected that the cluster size
distribution could asymptotically approach a pure power-law behaviour as temperature further
decreases and approaches some critical point if possible.
The similarity of the clusters size distributions at different temperatures is reminiscent of a general scaling behaviour.
Therefore, to explore the scaling law behind the cluster size distributions, a general scaling ansatz\cite{Christensen-book}
for the cluster number density $n(S,T)$ based on percolation theory was employed. That is,
\begin{equation}\label{ansatz1}
n(S,T) \propto S^{-\tau}f(S/S_c),
\end{equation}
where the characteristic cluster size $S_c$ diverges as a power-law in terms of the distance of $T$ from $T_c$:
\begin{equation}\label{ansatz2}
S_c\propto\left|1-T_c/T\right|^{-1/\sigma},
\end{equation}
and the function $f$ is known as the scaling function for the cluster number density.
Generally, the expression of $f$ varies from system to system, and dimension to dimension.
Analytic solution of $f$ is non-trivial in most cases. However, the asymptotic behaviour
of $f$ can provide sufficient information of the scaling behaviour of the size distributions.
According to percolation theory,
for $S/S_c\ll1$, the scaling function is approximately constant, and for $S/S_c\gg1$ it decays rapidly.
To our knowledge\cite{Christensen-book}, except for one-dimension percolation, this behaviour of $f$ is
quite universal in the scaling analysis based on above scaling ansatz.

To reveal the underlying scaling behaviour of the cluster size distributions, following above
scaling ansatz, we scaled the cluster number density $n(S,T)$ with $S^{-\tau}$ and the cluster size $S$
with  $\left|1-T_c/T\right|^{-1/\sigma}$, respectively. Figure \ref{sl} shows that
the scaled cluster size distributions at different temperatures collapse onto
a master curve representing the graph of scaling function $f(S/S_c)$. Here $T_c$=700 K is the scaling
crossover-temperature. $\tau$=2.05 and $\sigma$=0.4 are the scaling exponents.
These two exponents are different from the values obtained in three-dimensional site
percolation\cite{Lorenz-pre,Tiggemann-ijmpc,Ballesteros-jpa},
because the ISROs percolate in a continuous space, but not on a discrete periodic lattice.
As shown in Fig.~\ref{sl}, the scaling function $f(S/S_c)$ decays rapidly as the scaled
cluster size $S_T\gg1$, which indicates that there is a characteristic
cluster size in the system. The evolution of the characteristic cluster size $S_c$ shows a divergence
behaviour as temperature is approaching the critical point $T_c$. At $T=T_c$,
the characteristic cluster becomes infinite, so that $n(S,T) \propto S^{-\tau}f(S/S_c)$ becomes
\begin{equation}\label{ansatc}
n(S,T)\propto S^{-\tau},
\end{equation}
because it can be seen from Fig.~\ref{sl} that the scaling function $f(S/S_c)$ will
approach a non-zero constant for $S_T\ll1$. The scale-free behaviour of the
cluster size distribution at threshold shows some information about the
geometric properties of the percolating cluster, indicating that the
incipient infinite cluster has an internal fractal geometry. The fractal dimension
$d_f\sim2.86$ of the structures formed by connected ISROs at the critical
point can be calculated from the scaling relation of
\begin{equation}\label{fd}
\tau=d/d_f+1,
\end{equation}
where $d=3$ is the spatial dimension\cite{Christensen-book}. It can be seen that the value of $d_f$ is
not that small, which indicates that the atomic packing of incipient infinite cluster is not so loose.
On the other hand, just as the characteristic cluster size $S_c$ diverges
as $T$ is approaching $T_c$, the correlation length associated with the
connected ISROs also diverges. For a particular characteristic cluster size $S_c$,
the associated radius of gyration defines a characteristic length scale that is
proportional to the correlation length $\xi$. According to percolation theory, one has
\begin{equation}\label{cs}
S_c\propto\xi^{d_f},
\end{equation}
so that the temperature dependence (as $T\rightarrow T_c$) of the correlation length
$\xi$ can be characterized as
\begin{equation}\label{cl}
\xi\propto\left|1-T_c/T\right|^{-\nu},
\end{equation}
which shows a power-law behaviour with a divergence at $T=T_c$.
The corresponding critical exponent $\nu=0.875$ can be determined by the scaling relation\cite{Christensen-book} of
\begin{equation}\label{ce}
\nu=1/\sigma d_f=(\tau-1)/\sigma d.
\end{equation}

As shown above, in percolation, once the cluster number density is known, all other
quantities can be derived. All results indicate that the percolation of ISROs forming the so-called
icosahedral medium-range orders indeed correlates with the glass transition
in CuZr metallic glass-forming liquids. We argue that $d_f$ of the incipient infinite cluster
emerging at $T_c$ cannot be calculated directly by the box-counting method. The reason is as follows:
Since the crossover temperature $T_c$ is blow the glass transition
temperature (over 900 K) obtained in the previous study\cite{Mendelev-pm},
the atomic structures at 700 K obtained in simulations are in non-equilibrium states and sensitive
to the cooling-rate. As a result, $d_f$ obtained from box-counting analysis for the atomic
structures at 700 K will be cooling-rate dependent. This is not exactly the same $d_f$  
derived in percolation theory for the equilibrium structural phase transition.
Therefore, $d_f$ obtained from the universal scaling relation ($\tau=d/d_f+1$)
is generic. It is also worth noting that $T_c$ is not cooling-rate dependent, because
it is a critical temperature derived from the equilibrium structural phase transition of the modelling system, and
all data for determining $T_c$ are generated from the relatively equilibrium
supercooled liquid states. This is similar to T$_0$ in
Vogel-Fulcher-Tamman (VFT)\cite{Debenedetti-nat,Berthier-rmp} equation.

\subsection*{Physical origin of dynamic slowdown associated with percolation}

\begin{figure}[ht]
\centering
\includegraphics[width=0.5\linewidth]{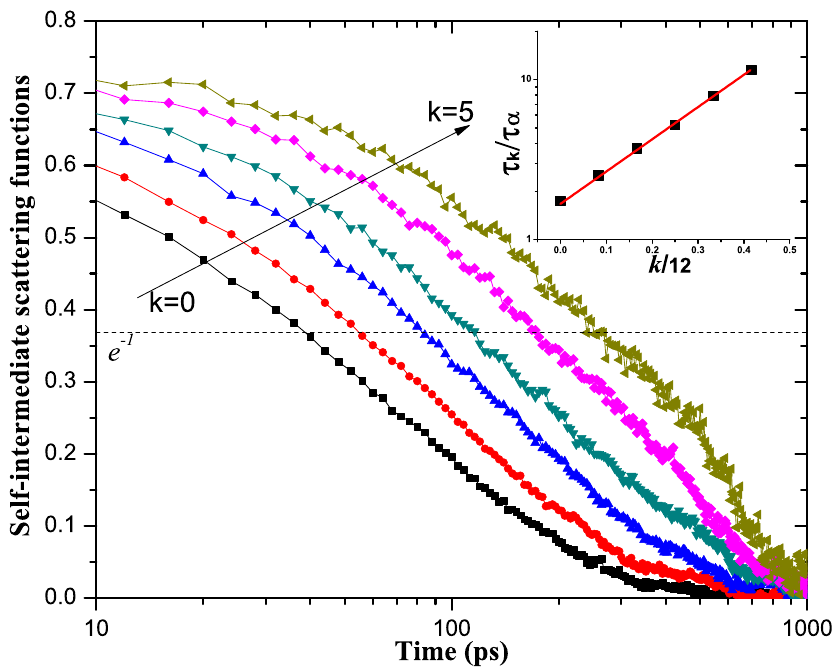}
\caption{The self-intermediate scattering functions of the central atoms of ISROs with
different node degree $k$ in CuZr metallic glass-forming liquid at 1000 K.
The inset shows an exponential dependence of scaled relaxation times
$\tau_k/\tau_{\alpha}$ on $k/12$.\label{isf}}
\end{figure}

The above scaling analysis demonstrates that the percolation of ISROs in CuZr
metallic glass former during cooling is closely related to glass transition.
It is still not clear why the percolation of ISROs could contribute to slowing down of dynamics.
As mentioned above, the connectivity of ISROs significantly influences the dynamical property
of local structures\cite{Wu-prb}. In order to emphasize the importance of the medium-range
structures formed by the connection of ISROs to the dynamic slowdown, we calculated the
self-intermediate scattering functions\cite{Kob-pre} (SISFs) of the central atoms of icosahedra
with different node degree
\begin{equation}
F_s^k(q,t)=\frac{1}{N_k}\left\langle\sum_{j=1}^{N_k}\exp\{i\vec{q}\cdot[\vec{r}_j(t)-\vec{r}_j(0)]\}\right\rangle,
\end{equation}
where the sum is over all central atoms with degree $k$, $\vec{r}_j(t)$ is the location of
atom $j$ at time $t$. $\vec{q}$ is chosen with the amplitude approximately equal to the
value of the first peak position in the static structure factor, and $\langle\cdot\rangle$ denotes
the ensemble average. The relaxation time $\tau_k$ is determined by $F_s^k(q,\tau_k)=e^{-1}$.
Figure \ref{isf} shows the SISFs as a function of time for central atoms of the icosahedra with
different $k$ values (due to the very rare connection of $k>5$, SISFs of $k>5$
were not calculated\cite{Wu-prb}.) in supercooled metallic glass-forming liquid at 1000 K.
It can be seen that all SISFs for different $k$ values exhibit non-exponential decay behaviour,
and the decay becomes much slower as $k$ value increases, dramatically
depending on $k$. The inset in Fig.~\ref{isf} explicitly shows that the scaled relaxation time
$\tau_k/\tau_{\alpha}$ increases exponentially with the value of $k/12$ ($\tau_{\alpha}$
is the average structural relaxation time, and 12 is the maximum value of node degree)\cite{Wu-prb}.
It can be seen that the relaxation time of the atoms with large $k$ is even more than 10 times of the $\tau_{\alpha}$.
This finding reveals that the medium-range structures formed by the connection of ISROs fundamentally
influences the relaxation dynamics of local atomic structures. As the icosahedral medium-range order
percolates during quenching, the resulting dynamic slowdown spreads
in the whole system, leading to the sluggish dynamics, which finally contributes to glass transition.

\begin{figure}[ht]
\centering
\includegraphics[width=0.5\linewidth]{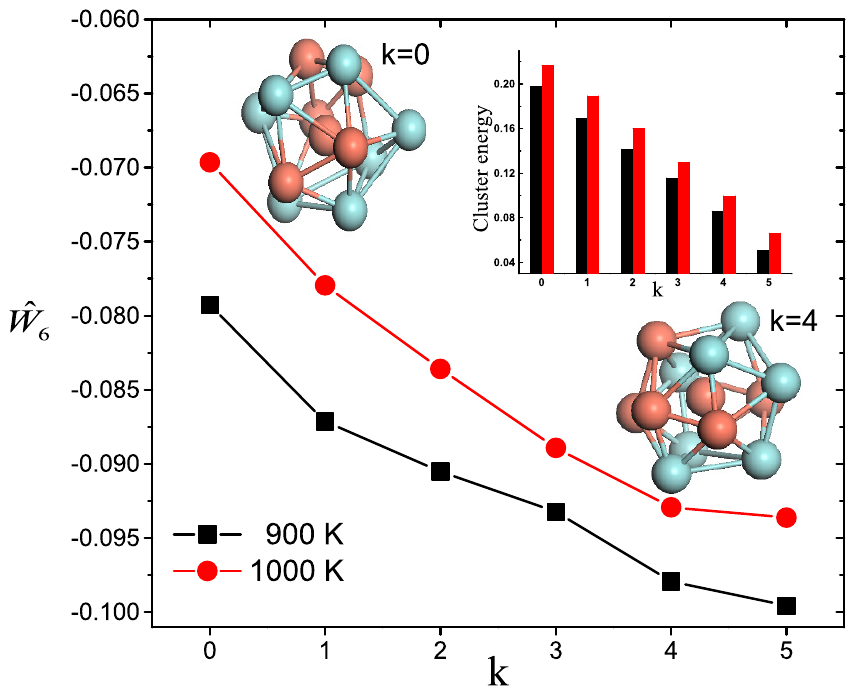}
\caption{The dependence of $\hat{W}_6$ of an ISRO on its node degree $k$ in supercooled liquids
at 900 K (black) and 1000 K (red), respectively. The inset histogram shows the change of the
formation energies of ISROs with different node degree $k$. Two atomic configurations of ISROs
with $k=0$ and $k=4$ obtained from MD simulation were also presented for geometric comparison.\label{we}}
\end{figure}

It is also very interesting to investigate how the icosahedral medium-range order influences
the atomic symmetry of icosahedra themselves. All icosahedra in the MD
modeled samples are actually distorted from the ideal icosahedron, because from
a geometrical viewpoint, icosahedral clusters cannot fill the entire three-dimensional space
without partially breaking of the five-fold rotational symmetry\cite{Hirata-sci}.  As observed
by Frank over half a century ago, the ideal icosahedral arrangement indeed has
a significantly lower energy than fcc atomic arrangement\cite{Frank-prsl}. The question is whether the connection of ISROs will promote
dense packing and five-fold local symmetry of ISROs. If so,
the self-aggregation effect of icosahedra\cite{Li-prb} will naturally tend to minimize the local energy density,
slow the atomic dynamics, and lead to great geometric frustration. In order to get deep insight
into the above discussion, we analyzed the local atomic symmetry of ISROs and its dependence on
the connectivity degree $k$. To analyze the local atomic symmetry of ISROs,
the bond orientational order (BOO) parameter introduced by
Steinhardt \emph{et al}\cite{Steinhardt-prb} was adopted, in which the BOO of the $\ell$-fold symmetry is defined as a $2\ell+1$ vector:
\begin{equation}\label{boo}
q_{\ell m}(i)=\frac{1}{N_i}\sum Y_{\ell m}(\theta(\vec{r}),\varphi(\vec{r})),
\end{equation}
where $Y_{\ell m}$ is spherical harmonics and $N_i$ is the number of bonds of atom $i$ with its
nearest neighbor atoms. In the analysis, one use the rotational invariants defined as
\begin{equation}\label{ql}
q_{\ell}=\left(\frac{4\pi}{2\ell+1}\sum_{m=-\ell}^{\ell}|q_{\ell m}|^2\right)^{1/2},
\end{equation}
\begin{equation}\label{wl}
W_{\ell}=\sum_{m_1+m_2+m_3=0}\left(\begin{array}{ccc} \ell & \ell & \ell \\ m_1 & m_2 & m_3 \end{array}\right)q_{\ell m_1}q_{\ell m_2}q_{\ell m_3},
\end{equation}
\begin{equation}\label{wh}
\hat{W}_{\ell}=W_{\ell}\left(\sum_{m=-\ell}^{\ell}|q_{\ell m}|^2\right)^{-3/2},
\end{equation}
where the term in brackets in the invariants (\ref{wl}) is the Wigner $3j$ symbol\cite{Leocmach-nc}.
For fcc, bcc, and hcp symmetry, the value of $\hat{W}_6$ is close to zero, while for perfect icosahedron,
$\hat{W}_6=-0.169757$, so that $\hat{W}_6$ is sensitive to reflect the five-fold atomic symmetry features of ISROs
in metallic glass-forming liquids. Figure \ref{we} shows that the average values of
$\hat{W}_6$ of ISROs are not -0.169757, but in between -0.06 and -0.10, suggesting that the
ISROs in metallic glass-forming liquids are not perfect, but distorted with partially fcc symmetry,
in agreement with the previous studies\cite{Hirata-sci}. Furthermore, as shown in Fig.~\ref{we},
with increasing node degree $k$, $\hat{W}_6$ decreases to more negative values.
This indicates that as the node degree increases, on the range we studied, the atomic configurations of ISROs
are optimized towards more regular icosahedra.
The configuration optimization of ISROs produces significant geometric
frustration in the supercooled liquids.

We also calculated the formation energy $E_{p,c}$ of ISROs in the supercooled metallic liquids
and investigated the dependence of the formation energy on node degree $k$. The formation energy
of ISROs can be calculated by the formula of
\begin{equation}\label{fe}
E_{p,c}=\frac{1}{N}\sum_j\left(E_{p,j}-E_{ref,a}\right),
\end{equation}
where $E_{p,j}$ is the potential energy of the $j$th atom in an icosahedron
and $E_{ref,a}$ is the reference energy of the element $a$\cite{Wu-apl}.
In our calculations, the chemical potentials of Cu and Zr in crystal structures (fcc for Cu and hcp for Zr) were
used as the reference energies for Cu an Zr atoms, respectively. Therefore, in equation (\ref{fe}),
the effect of the chemical compositions in an icosahedron was eliminated, and the formation energies
of ISROs with different chemical compositions can be comparable. Thus, we can analyze the
formation energies of ISROs with different degree $k$. The inset histogram in Fig.~\ref{we} clearly shows
that with increasing degree $k$, the formation energy of ISROs decreases monotonically,
indicating that the formation of the icosahedral medium-range order leads to lower energies,
and the structures become more stable. All of above results indicate that the connection
between different icosahedra will indeed generate a positive feedback for optimization of
the local ISROs. This effect will minimize the local energy density, slow down the atomic dynamics,
cause great geometric frustration, and finally contributes to glass transition.

In summary, we carried out MD simulations for CuZr metallic glass-forming liquids
to investigate the relationship between percolation of ISROs and glass transition.
As the system is supercooled, the ISROs tend to connect with each other to form big clusters.
The cluster size distributions at different temperatures follows an excellent scaling law,
which indicates that the cluster size distribution evolves toward a power-law behaviour,
and an infinite size cluster is formed as the system is approaching a critical point.
Therefore, there would be a structural phase transition manifested by percolation
of locally favoured structures underlying the glass transition if the liquid could be cooled
slowly enough but without crystallization intervening.
Furthermore, the percolation of ISROs and the formation of medium-range orders
optimize the atomic configurations of ISROs towards more regular icosahedra,
enhancing the geometric frustration and minimizing the local energy density, which leads
to the dynamic slowdown of the metallic glass-forming liquids. Our findings suggest that the
geometric phase transition manifested by percolation of locally favoured structures
could be critical for the understanding of the nature of glass transitions.

\section*{Methods}

\subsection*{Molecular dynamic simulations}
In this work, molecular dynamic (MD) simulations were carried out for Cu$_{50}$Zr$_{50}$
metallic alloy using the LAMMPS package\cite{Plimpton-jcop}. The interatomic interaction
was described by the well-developed embedded-atom method potential for CuZr alloys\cite{Mendelev-jap}.
The sample contains 10000 atoms being randomly distributed in a cubic box with periodic boundary condition applied
in three dimensions. The MD step is 2 fs. At first, the sample was equilibrated at 2000 K
for 4 ns (2,000,000 MD steps) in NPT (P=0) ensemble with Nose-Hoover
thermostat and barostat. The liquid was then quenched at a rate of 1 K/ps down
to its target temperature (1600K, 1500 K, 1400 K, 1300 K, 1200 K, 1100 K, and 1000 K, respectively).
During cooling, the box size was adjusted to maintain zero pressure.
At each temperature, the atomic configuration was relaxed in NPT (P=0)
ensemble for another 2 ns (1,000,000 MD steps) for the analysis of physical properties
(500 atomic configurations were collected for ensemble average). The structural
relaxation time ($\tau_{\alpha}$) of our modelling system at 1000 K was order
of magnitude of 10 ps, so the ensemble average window for analysis was
much longer than $\tau_{\alpha}$ of the system at each temperature we studied.
Note that at temperatures higher than the glass transition point,
the structures corresponding to (metastable) equilibrium can be achieved quite rapidly.
Therefore, in general, the effect of cooling rate on the analysis of structural, dynamic, or
other physical properties of our modelling system is eliminated.

\subsection*{Icosahedral network and node degree}
The local atomic structures in supercooled liquid samples at different temperatures
were analyzed by the Voronoi tessellation method\cite{Finney-prsc,Finney-nat,Borodin-pma}
and identified in terms of the Voronoi index $\langle n_3,n_4,n_5,n_6 \rangle$,
where $n_i$($i$=3,4,5,6) denotes the number of $i$-edged faces of a Voronoi polyhedron.
In our analysis, a cutoff distance of 5 \r{A} was chosen so that the Voronoi index distribution
was converged. To characterize the connectivity of ISROs in the system, we introduced
the \emph{graph theory}\cite{Newman-siam,Newman-pnas,Albert-rmp}.
In our scheme\cite{Wu-prb,Wu-jcp}, the central atom (with Voronoi index $\langle 0,0,12,0 \rangle$)
of an icosahedron is treated as a node, and two nodes are considered to be connected
if they are the nearest neighbors with each other, that is, the two icosahedra are interpenetrating or volume-sharing.
The choice of connection criterion is reasonable based on recent experimental observation\cite{Liu-prl},
in which the extent of icosahedral short-range order to form medium-range order is
consistent with a facing-sharing or interpenetrating configurations. Therefore,
the case of interpenetrating configuration was considered in our analysis.
Property variation depending on the connection criterion will be discussed in detail in future works.
With the definitions of nodes and edges abstracted from the atomic modelling system,
we established the icosahedral network. Based on the scenario of graph theory, the node
degree $k$ was defined as the number of other nodes directly connected to it.
The maximum value of node degree for the central atom of an icosahedron is 12.
Our results suggest that node degree introduced from graph theory is a
good nonlocal (to some extent) order parameter for classifying atoms with distinct properties.

\section*{Acknowledgements}

This work was supported by NSF of China (Nos. 51271197 and 51271195), the MOST Project of China (No. 2015CB856800 and 2012CB932704).

\section*{Author contributions}

Z.W.W. and M.Z.L. conceived and designed the research and analysis. Z.W.W., F.X.L., and C.W.H. performed the molecular dynamics simulations. Z.W.W. and M.Z.L. wrote the paper. All authors discussed the results and revised the manuscript.

\section*{Additional information}

\textbf{Competing financial interests:} The authors declare no competing financial interests.

\end{document}